# Carbon dioxide emission during forest fires ignited by lightning

Magdalena Pelc[*]

Physics Institute, Maria Curie – Sklodowska University

Lublin, Poland

Radoslaw Osuch

High School of Ecology and Management

Warsaw, Poland

**Abstract**

In this paper we developed the model for the carbon dioxide emission from forest fire. The master equation for the spreading of the carbon dioxide to atmosphere is the hyperbolic diffusion equation. In the paper we study forest fire ignited by lightning. In that case the forest fire has the well defined front which propagates with finite velocity.

[*] corresponding author: magdalenapelc@o2.pl

## 1. Introduction

An important problem in our society is the forest fire. Some important organizations, especially the USDA Forest Service, have been researching this theme for some time.
On the other hand recently several papers related with percolation theory and self-organized criticality (SOC) [1] are trying to provide a different dynamical model for the fire spread. The basic problem of the SOC models in their hard adaptation to the real problems.

The model we present in this paper is more realistic than the SOC models. We describe the forest fire as the process with well defined memory function. To that aim we restricted our description to the lightning induced forest fire. In that case we have well established time scales, for the lightning duration is much shorter than the characteristic time of the fire spreading.

In our paper the signature of the forest fire is the emission of $CO_2$. For the study of the carbon dioxide emission we developed and solved the hyperbolic diffusion equation. As the result history of the emission process can be established. The obtained solution of the hyperbolic equation is not singular at the time $t = 0$ and gives the finite value of the speed of the fire front.

## 2. The model

Fick diffusion equation is a special case of the parabolic transport equation in which speed of perturbation is infinite. Parabolic transport equation has been applied to the dispersion of biological population [2], epidemic models [3] and chemical systems [4].

However, if the thermal interval of the perturbation of the system is very short in comparison to the characteristic time of the system, Fick law does not hold and the diffusion equation – parabolic – must be generalized to the hyperbolic diffusion equation [5].

The lightning pulse energy duration is of the order of $10^{-6}$ s = 1 μs and the characteristic time for fire propagation in forest is of the order of seconds [6]. The forest wood combustion produces $CO_2$ in two steps. First step with time interval of the 1 μs in which the hyperbolicity of the fire transport equation is important. The second step is the thermal processes which can be described by parabolic diffusion equation.

Let us consider the master equation for the $CO_2$ density $n(x,t)$. To that aim we formulate the generalized Fourier law [5]

$$\vec{q} = -\int_{-\infty}^{t} K(t-t')\nabla n(x,t')dt', \qquad (1)$$

where $q(x,t)$ is the density of a $CO_2$ flux and $K(t-t')$ is a memory function for combustion process. The density of $CO_2$ flux satisfies the following equation of $CO_2$ emission [5]:

$$\frac{\partial}{\partial t}n(x,t) = const\int_{-\infty}^{t} K(t-t')\nabla n(x,t')dt'. \qquad (2)$$

We introduce the following equation for the memory function describing the combustion processes

$$K(t-t') = const \lim_{t_0 \to 0}\delta(t-t'-t_0). \qquad (3)$$

In that case the combustion processes have a very short memory. Combining Eqs. (3) and (2) we obtain:

$$\frac{\partial}{\partial t}n(x,t) = const\nabla^2 n(x,t). \qquad (4)$$

Equation (4) is the Fick equation for the emission of $CO_2$.

For a system with a long memory the memory function has the form

$$K(t-t') = const. \qquad (5)$$

In that case the master equation for the combustion process with emission of $CO_2$ is:

$$\frac{\partial^2 n(x,t)}{\partial t^2} = const\nabla^2 n. \qquad (6)$$

Equation (6) is the wave equation. Considering the standard form the wave equation we put $v^2 =$ const. and obtain

$$\frac{1}{v^2}\frac{\partial^2 n(x,t)}{\partial t^2} = \nabla^2 n, \qquad (7)$$

where $v$ is the speed of combustion processes.

In realistic combustion processes the memory function is intermediate and has the form

$$K(t-t') = const \exp\left[-\frac{(t-t')}{\tau}\right]. \qquad (8)$$

The parameter $\tau$ is the relaxation time for combustion processes. In that case the master equation for combustion processes is [5]

$$\frac{\partial^2 n}{\partial t^2} + \frac{1}{\tau}\frac{\partial n}{\partial t} = v^2\nabla^2 n. \qquad (9)$$

For very short lightning pulse we put in (9)

$$n(x,t) = \exp\left[-\frac{1}{2\tau}(t-t_0)\right] f(x,t) \qquad (10)$$

and we obtain

$$\tau \frac{\partial^2 f(x,t)}{\partial t^2} - v^2\tau \frac{\partial^2 f(x,t)}{\partial x^2} - \frac{1}{4\tau} f(x,t) = 0. \qquad (11)$$

The solution of (11) has the form [4]

$$f(x,t) = \frac{1}{\sqrt{4v^2\tau^2}} I_0\left[\frac{1}{\sqrt{4v^2\tau^2}} \sqrt{v^2(t-t_0)^2 - (x-x_0)^2}\right]$$
$$\qquad \text{for } |x-x_0| < v(t-t_0) \qquad (12)$$
$$f(x,t) = 0 \qquad \text{for } |x-x_0| > v(t-t_0)$$

In Eq.(12) $I_0(x,t)$ is the modified Bessel function; $I_0$ has the asymptotic form [5]

$$I_0(z) = \frac{1}{\sqrt{2\pi z}}, \qquad z \to \infty. \qquad (13)$$

Substituting formula (13) to (12) and (11) we obtain in the limit $\tau \to 0$ (Fick approximation)

$$n(x,t) = \frac{1}{\sqrt{4\pi v^2\tau(t-t_0)}} \exp\left[-\frac{(x-x_0)^2}{4v^2\tau(t-t_0)}\right] \qquad \text{for } t > t_0$$
$$n(x,t) = 0 \qquad \text{for } t < t_0 \qquad (14)$$

Formula (14) is the well known solution of the diffusion equation for Dirac type initial condition. In that case $v^2\tau = D$, where $D$ is the diffusion coefficient for the emission of $CO_2$

3. **Conclusion**

In this paper we developed and solved the hyperbolic diffusion equation for the $CO_2$ emission ignited by the forest fire. In the paper we restricted our analysis to the forest fire triggered by lightning. In USA the lightning ignited 10 % of the total number of forest fire.